\documentstyle[aps,epsfig,twocolumn]{revtex}
\begin{document}

\title{Photoluminescence of a single InAs quantum dot molecule\\
       under applied electric field}
\draft
\author{I. Shtrichman, C. Metzner, B. D. Gerardot, W. V. Schoenfeld, and P. M. Petroff}
\address{Materials Department, University of California, Santa Barbara, CA 93106, USA}
\date{\today}
\maketitle

\begin{abstract}
We study the electronic coupling between two vertically stacked InAs quantum dots,
which are embedded in the center of a n-i-n structure. We use a
micro-photoluminescence setup to optically isolate a single quantum dot pair and
measure the time-averaged photoluminescence under an applied vertical electric field. We
find that field tunable coupling between excited states of the two quantum dots leads
to charge transfer from one dot to the other. We model the spectra including
simultaneously the field dependent charge transfer and exciton capture rates, and the
many-body spectra of the quantum dot molecule for different carrier configurations.
\end{abstract}
\pacs{PACS: 73.21.La, 78.55.Cr, 78.67.Hc}

The emerging field of quantum computation has attracted great interest over the last
few years \cite{ref1}. Various theoretical schemes were proposed for the
implementation of quantum bits (qubits) and quantum gates, using semiconductor quantum
dots (QDs) \cite{ref2}. Specifically, the vertically stacked double QD system was
suggested to host a single \cite{ref3,ref4} or two qubits \cite{ref5}. One can then
control the coherent two-level system (qubit) with short optical pulses
\cite{ref3,ref5}, by an applied electric field \cite{ref4,ref6,ref7}, or by a magnetic
field \cite{ref6}. Once the basic quantum operation in such a system is achieved,
scaling up to high-density self-assembled ordered arrays of these units should be
feasible \cite{ref8}. A necessary step towards realization of a single qubit in a QD
pair is to achieve electronic (wavefunction) coupling between the two dots. In recent
years, several attempts have been made in this direction by comparing samples with 
different inter-dot spacing \cite{ref9,ref10}.

Clearly, the coupling between the two QDs in the pair is highly sensitive to their
relative energy levels. These energies are fixed for each QD by its dimensions and
material composition, which are hard to control, especially in the technologically
important type of self assembled QDs. However, by varying an electric field across the
QD pair one can tune the electronic states of the two QDs into and out of resonance.
This method allows one to investigate the electronic coupling between the two dots in a 
single, specific QD molecule, thus avoiding the difficulty of comparing different 
molecules from various samples with each other.

In this work we study the photoluminescence (PL) spectra of two vertically stacked QDs
as a function of excitation intensity and external electric field. We compare the
spectra of a single dot (QD atom), an electronically uncoupled QD pair and a coupled
QD molecule. For the QD molecule we find that the two dots have a large ground state
energy difference and that coupling occurs between their excited states. By tuning the 
electric field across the molecule we control the transfer of charge between the two dots, 
which is revealed in the time-averaged emission spectra, in similarity to recent works 
on charged single QDs \cite{ref11}.

Three sample structures were grown by molecular beam epitaxy (MBE) on GaAs substrates.
The samples contain an intrinsic layer embedded between two n-doped GaAs layers (100
nm thickness, $1\!\cdot\!10^{19} \mbox{cm}^{-3}$ Si dopant density), which serve as the
front and back electrodes of the device (n-i-n structure). The intrinsic layer
contains either a single InAs QD layer (sample A), or two vertically stacked strain
coupled QD layers separated by a GaAs spacer of d $\!=\!15$ nm (sample B) or d $\!=\!4.5$ 
nm (sample C). The QD layers are located in the center of a thick GaAs layer (200 nm), 
which in turn is surrounded by two superlattice barriers (20 periods of $0.5/0.5$ nm 
GaAs/AlAs) to reduce current flow through the device.

The self-assembled QDs were grown using the partially covered island technique \cite{ref12}. 
The lens shaped QDs in the first grown layer have a height of $\approx\!3$ nm and a
lateral size of $\approx\!50$ nm. Due to the strain field from the first QDs layer,
the dots in the second InAs layer tend to nucleate directly above the dots in the
first layer \cite{ref13,ref14}. For a QD layer separation of 15 nm it was shown that
the pairing probability is higher than $0.9$, while for $4.5$ nm separation the
probability is close to 1 \cite{ref13}. The growth of the second layer QDs is strongly 
influenced by the strain from the first layer, resulting large size and composition 
difference between the two dots in the pairs \cite{ref13,ref14,ref14a}.

The samples were not rotated during the InAs deposition and therefore a gradient in 
the QD density was formed across the wafers. A low QD density part of each of the
three samples was processed by conventional lithography. A shallow mesa was 
chemically etched to define the device area and isolate the top conducting n-doped GaAs 
layer. Next, a semi-transparent 6 nm thick Pt layer was evaporated on the surface to 
provide the top gate. A $400\,\mu$m square gold frame was then deposited on top of the
Pt layer to serve as the front contact to the sample. An ohmic contact to the back
electrode was accomplished by deposition and rapid thermal anneal of an AuGeNi alloy. 
We note that the devices are rectifying at low temperatures and in the relevant range of 
applied voltages the current is well below 100 $\mu$A.

For PL measurements of the high QD density parts of the wafers 
($\approx\!10^{10} \mbox{cm}^{-2}$) we used a closed cycle Helium cryostat to cool down
the samples to a temperature of 3 K. A HeNe laser was used for the excitation, and the
PL signal was collected and dispersed by a 0.19 m single monochromator followed by a
liquid nitrogen cooled CCD detector. The spectral resolution was limited to $0.6$ meV
(FWHM). For micro-PL measurements of the low QD density parts of the wafers
($\approx\!5\!\cdot10^6 \mbox{cm}^{-2}$) we used a Helium flow cryostat to cool down
the samples to 15 K. An objective lens, used for both tight-focus excitation and
luminescence collection, was located inside the cryostat, enabling a short working
distance from the sample surface (1.3 mm). In order to achieve high spatial resolution 
for the collected PL signal in addition to the excitation, a $1\,\mu$m object was defined 
on the sample by imaging it on a $50\,\mu$m pinhole, thus filtering out any light coming 
from outside the object light cone. The spatially filtered luminescence was then dispersed 
and detected by the same setup as in the conventional PL. The overall spatial resolution 
was measured to be $1.1\,\mu$m (FWHM) at 843 nm, while the average distance between the 
dots was about $5\,\mu$m, enabling us to optically isolate and study a single QD pair.

When an electric field is varied at fixed optical excitation intensity, the effective 
rate of exciton capture into the QDs changes drastically due to field-induced carrier
escape processes. For the micro-PL measurements under applied DC electric field we have 
significantly reduced this effect by using a continues wave Ti:Sapphire laser tuned to 
an energy below the GaAs barrier ($1.47$ eV), thus exciting only the InAs wetting layers 
(WLs).

We note that for each of the three samples we studied in detail at least 3 different
QD pairs (or dots in sample A), which show the same qualitative behavior for each
sample, although at slightly different energies.

\begin{figure}[!hbt]
\centering{\epsfig{figure=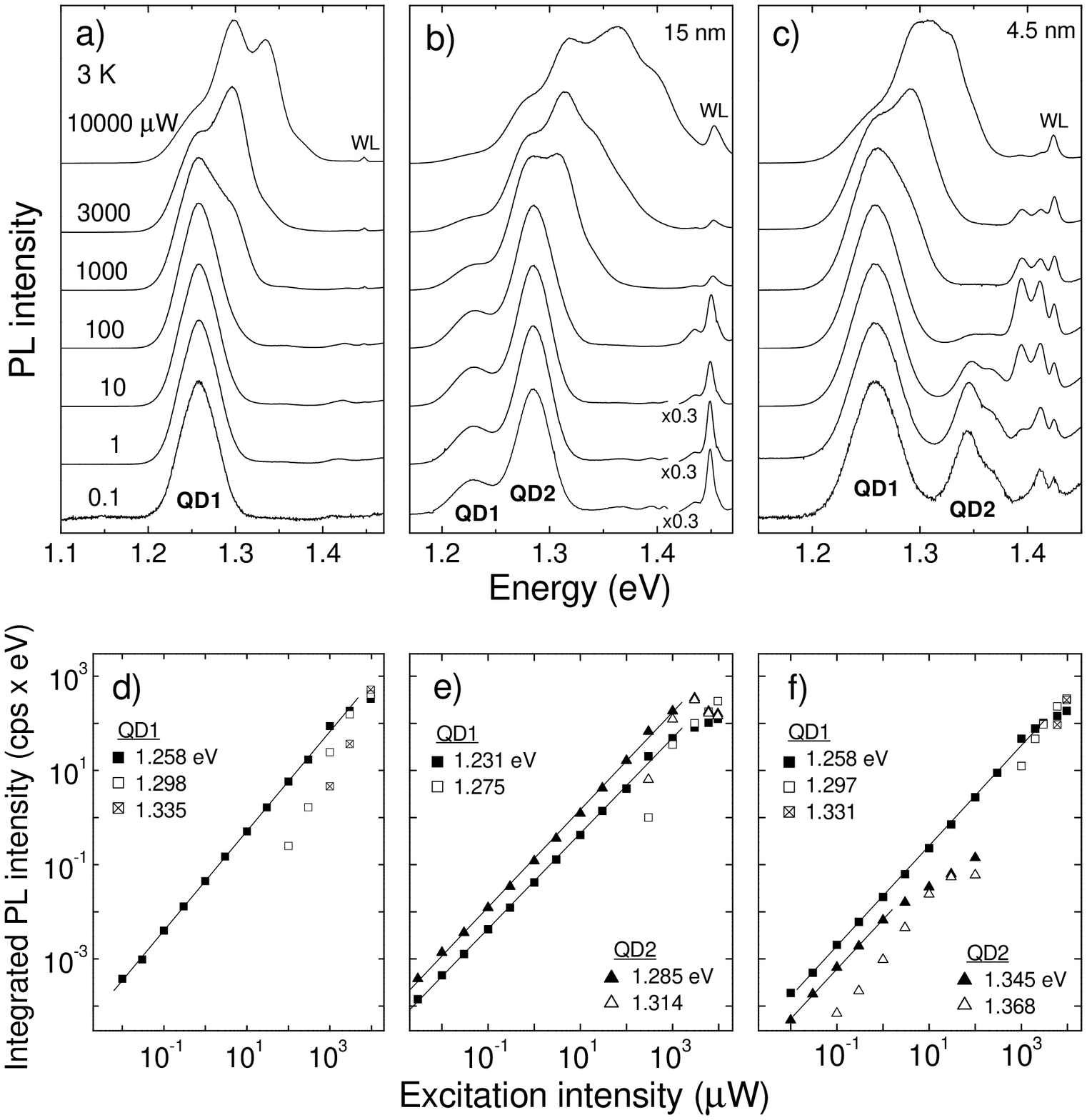,width=15cm}}
\caption{a-c) PL spectra from the high QD density part of samples A, B and C,
respectively, for different excitation intensities. The spectra are vertically
displaced for clarity. The excitation energy is 1.96 eV, and 1 $\mu$W corresponds to a
density of 0.02 W/$\mbox{cm}^2$. d-f) Spectrally integrated intensity of the different
PL peaks as a function of excitation intensity in samples A, B and C, respectively.
The solid lines are linear fits to the data.}
\label{fig1}
\end{figure}

In Fig.\ref{fig1} we display ensemble PL spectra of all three samples for different
excitation intensities, at zero applied voltage. Fig.\ref{fig1}a shows
the luminescence from a single layer of QDs (sample A), which at low excitation intensity
peaks at 1.258 eV and shows an inhomogeneous broadening of 50 meV (FWHM). At high
excitation intensities several excited states gradually appear in the spectra,
separated by $\approx\!40$ meV. The luminescence from the WL emerges at 1.447 eV. For
samples B (Fig.\ref{fig1}b) and C (Fig.\ref{fig1}c) we identify the two peaks, which
dominate the spectrum at the lowest excitation powers, as the inhomogeneously
broadened luminescence from the two QD layers. The low-energy peak (QD1) has about the
same energy as the single QD layer of sample A. We therefore relate this peak to the 
first grown (seed) layer of QDs in samples B and C. The high-energy peak (QD2) is assigned 
to the second, strain coupled layer of QDs. We attribute the separation of $\approx\!60$ meV 
(90 meV) between the QDs ground states in Fig.\ref{fig1}b (\ref{fig1}c) to the two dots
unintentional size and composition differences, and to the asymmetric strain field that 
they induce on each other. Several excited states of QD1 and QD2, separated by 40 meV
and 25 meV, respectively, are observed in the spectra as the excitation power
increases. We note that it is not necessary for the two QDs to have exactly the same ground 
state energies in order to achieve coupling between these states \cite{ref14b}. However, 
in our case, where the two ground states are separated by 90 meV (for d = 4.5 nm), the tunnel coupling between these states is negligible \cite{ref14c}.

\begin{figure}[!hbt]
\centering{\epsfig{figure=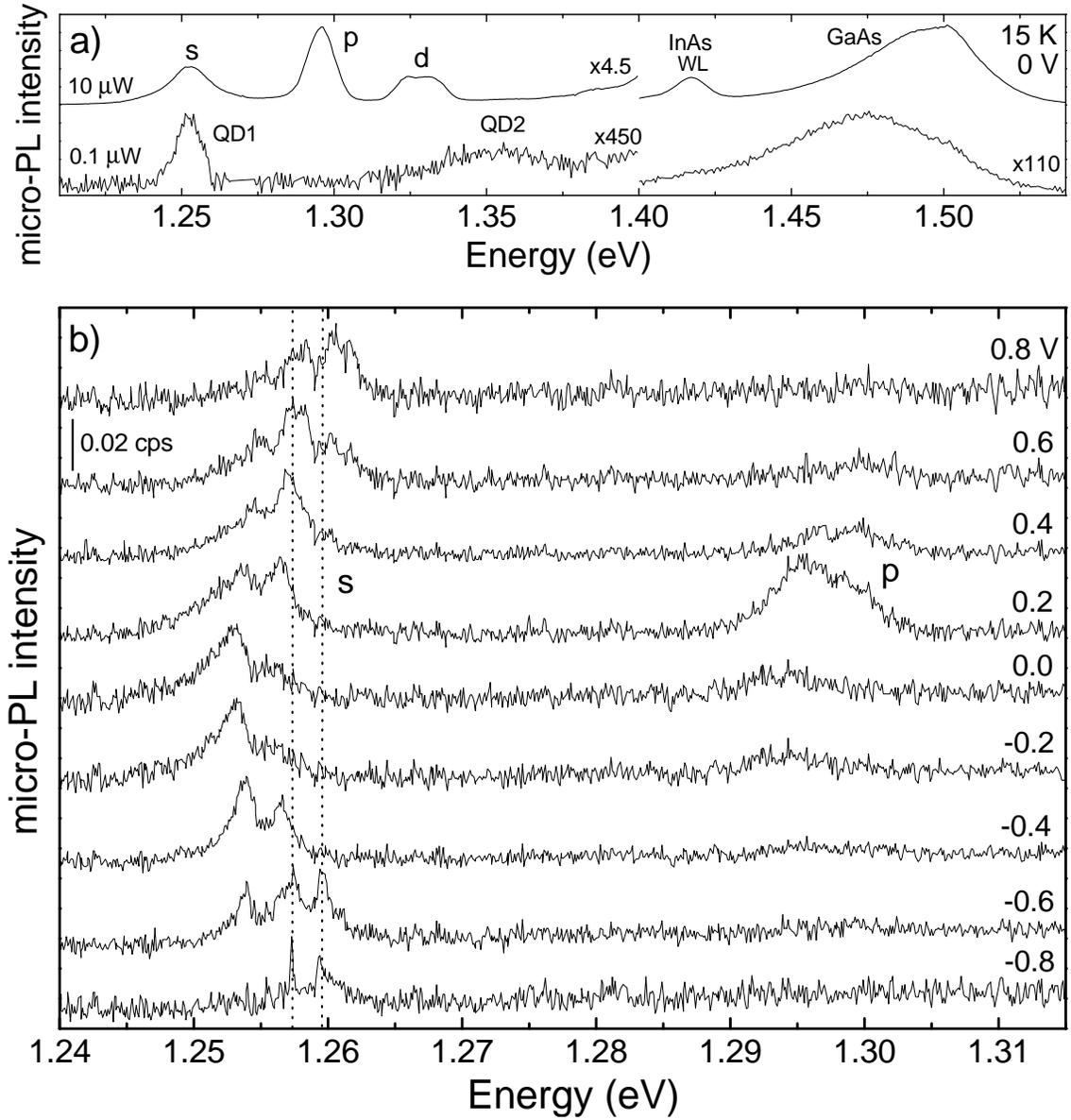,width=15cm}}
\caption{Micro-PL spectra of a single coupled QD molecule (sample C). a) Low 
resolution spectra excited at 1.96 eV and at zero applied voltage. b) High resolution 
spectra for different applied DC voltages. The excitation energy and intensity are 
1.47 eV and 100 $\mu$W, respectively, and 1 V corresponds to an electric field of 
38 kV/cm.}
\label{fig2}
\end{figure}

A clear difference between samples B and C is apparent from Fig.\ref{fig1}: In sample
B the PL intensity ratio of QD1 to QD2 is fixed at low excitation intensities
\cite{ref15}. In contrast, in sample C this ratio changes as the excitation
density increases. In Fig.\ref{fig1}d-f (samples A-C, respectively) we show the
spectrally integrated PL intensity of the different states of QD1 and QD2 as a
function of the excitation intensity. The various states are resolved using a
multi-Gaussian fit to each of the PL spectra of Fig.\ref{fig1}a-c. We find that in
all three samples the QD1 ensemble ground state PL grows linearly with the excitation
intensity over almost 6 orders of magnitude, due to gradual filling of the many dots in 
the ensemble. This is also the case for the QD2 ensemble ground state in sample B. In 
contrast, the ground state of QD2 in sample C saturates at much lower intensities, 
indicating that in this sample the two QDs are electronically coupled. Since QD2 
luminescence saturates while QD1 luminescence continues to grow linearly, we deduce 
that one type of charge carrier is transferred from QD2 to QD1, while the opposite 
charge remains in QD2 (thus forming an indirect exciton in the molecule). The fact that luminescence from the ground state of QD2 is apparent at low excitation intensities 
suggests that this charge transfer takes place between excited states of the two dots. 
We note that the qualitatively different behavior of QD1 and QD2 in the ensemble PL 
spectra of samples B and C is reproduced also in micro-PL measurements of single QD 
pairs (see for instance Fig.\ref{fig2}a, below).

In the micro-PL spectra of a single QD (sample A) we observe, as the excitation
intensity is raised, a number of sharp peaks in the spectra in agreement with previous
works \cite{ref16,ref17}. When varying the applied voltage at constant
photo-excitation intensity, we find that the peaks in the PL spectra change their
relative intensities due to a change in the effective exciton capture rate into the
dot (not shown). This effect is due to tunneling of carriers out of the InAs layer and
into the thick GaAs surroundings at high enough fields. We find that flat bands
condition, in which the capture rate into the InAs layer is maximized, is achieved at
0.2 V. This result is also in agreement with photocurrent measurements of the device.
Due to our limited spectral resolution we see no spectral shift related to the spatially 
direct Stark effect, which in the relevant range of electric fields is expected to be on 
the order of few hundreds $\mu$eV \cite{ref18}.

In the micro-PL spectrum of a single pair in sample B we observe the spectral signature 
of two independent QDs, which are separated by $\approx\!80$ meV (not shown). Under 
applied electric field the spectra from this electronically uncoupled QD pair strongly
resemble the single QD results.

This behavior is in contrast with the results from a QD molecule in sample C, as
displayed in Fig.\ref{fig2}. At zero applied voltage (Fig.\ref{fig2}a) and low
excitation intensity, QD1 and QD2 are clearly seen in the spectrum and are separated
by $\approx\!100$ meV. The luminescence from QD2 is broad, probably due to simultaneous recombination from the ground and excited states of this dot (compare to the ensemble 
case, Fig.\ref{fig1}c). At high excitation intensities QD2 disappears from the 
spectrum and the excited states of QD1 emerge gradually.

In Fig.\ref{fig2}b we show the high-resolution spectra of QD1 for different applied
voltages. We note that several peaks appear and disappear in the S shell luminescence
of QD1 while varying the electric field. By measuring the PL spectrum as a function of
the excitation intensity at various applied voltages (not shown), we identify the
transition energy of a single neutral exciton ($1X_S$) at 1.2595 eV, and of a neutral
bi-exciton ($2X_S$) at -2.1 meV relative to $1X_S$ (these transitions are marked by dotted
lines in Fig.\ref{fig2}b). In the -0.4 V spectrum we resolve additional two lines, at
-3.2 meV and -5.7 meV relative to $1X_S$. In the -0.2 V spectrum a shoulder appears at 
the low-energy part of the S shell spectrum.
We relate these lines and shoulder to recombination of an S shell exciton in the
presence of charged multi-exciton complexes. The above S shell picture then almost
symmetrically reverses at positive voltages. In addition to the S shell luminescence, a 
broad P shell peak emerges at $\approx\!1.296$ eV and is most intense at a voltage of 
0.2 V. As mentioned above, this behavior is due to flat bands condition at this voltage, 
in which the effective exciton capture rate of the QD molecule is maximized. We relate 
the broad P shell peak to luminescence from neutral and charged tri-exciton states, 
which are not resolved in this case.

\begin{figure}[!hbt]
\centering{\epsfig{figure=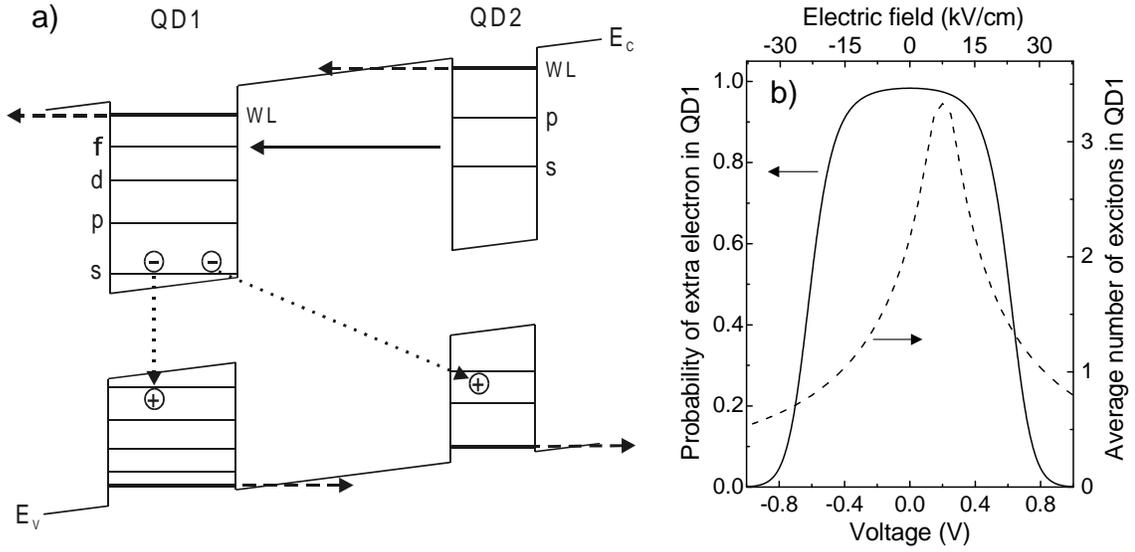,width=15cm}}
\caption{a) Schematic bandstructure along the growth direction of a QD molecule under
applied electric field. The arrows mark the main kinetic processes of our model. In
this example QD1 is populated by a single exciton plus one extra electron, which was
transferred from QD2, leaving behind a hole, i.e. $X_1\!=\!1,Y\!=\!1,X_2\!=\!0$. b)
Calculated probability of extra electron (solid line, left axis), and average number
$\overline{X_1}$ of excitons in QD1 (dash line, right axis) as a function of applied
voltage.}
\label{fig3}
\end{figure}

The evolution of the steady-state micro-PL spectra in Fig.\ref{fig2}b can be explained 
by a combination of two voltage-dependent effects: First, a change in the neutral exciton 
occupation numbers $X_1$ and $X_2$ in QD1 and QD2, respectively, due to a field-dependent 
escape rate (shown schematically in Fig.\ref{fig3}a). We estimate that the average exciton 
occupation $\overline{X_1}$ of QD1 drops from $\approx\!3$ at voltage U $\!=\!0.2$ V to 
$\approx\!1$ at U $\!=\!0.8$ V (dash line in Fig.\ref{fig3}b). Second, a change in the 
probability of single carrier transfer between the QDs, which has a broad maximum around 
U $\!=\!0$ V and becomes negligible at U $\!=\!\pm 0.8$ V (solid line in Fig.\ref{fig3}b). 
By comparing with theoretical PL-spectra (see below), calculated for different charging 
scenarios, we conclude that it is an electron rather than a hole that hops from QD2 to QD1. 
Whenever an electron transfer occurs, a spatially indirect electron-hole pair is formed in 
the QD molecule, and the resulting charge dipole prevents further transfer (Coulomb blockade 
effect). Thus, the number $Y$ of indirect electron-hole pairs is restricted to 0 or 1 in 
our case.

In order to support our hypothesis we compare the micro-PL spectra of Fig.\ref{fig2}b
to a numerical simulation, the details of which will be published elsewhere
\cite{ref19}. In our model we describe the confinement potential of the QD molecule in
the growth direction as an asymmetric double quantum well, and in the lateral directions 
as a rotationally symmetric parabolic well. Coulomb interaction effects are treated by
direct diagonalization (configuration interaction method \cite{ref20}) of the respective 
many-body Hamiltonian, which is defined by the momentary population $(\!X_1\!,\!Y\!,\!X_2\!)$ 
of the QD molecule. We assume complete energy and spin relaxation of the carriers prior
to each recombination event. Using the resulting correlated many-body states, we
compute the PL-emission spectrum $I(\hbar\omega,X_1,Y,X_2)$ for each relevant
population. The time-integrated spectrum $I_{av}(\hbar\omega,U)$ is calculated by
averaging $I(\hbar\omega,X_1,Y,X_2)$ over all possible populations, weighted by their
voltage-dependent formation probability $P(X_1,Y,X_2,U)$. This probability is determined
by the lifetimes of the exciton complexes, the escape rates of excitons from the WL
(modeled as a tunneling process out of a quantum well), and the effective rate of electron 
inter-dot transfer (modeled as an acoustic-phonon-assisted hopping process with
resonant tunneling enhancement and including level broadening due to the statistical
population fluctuations).

\begin{figure}[!hbt]
\centering{\epsfig{figure=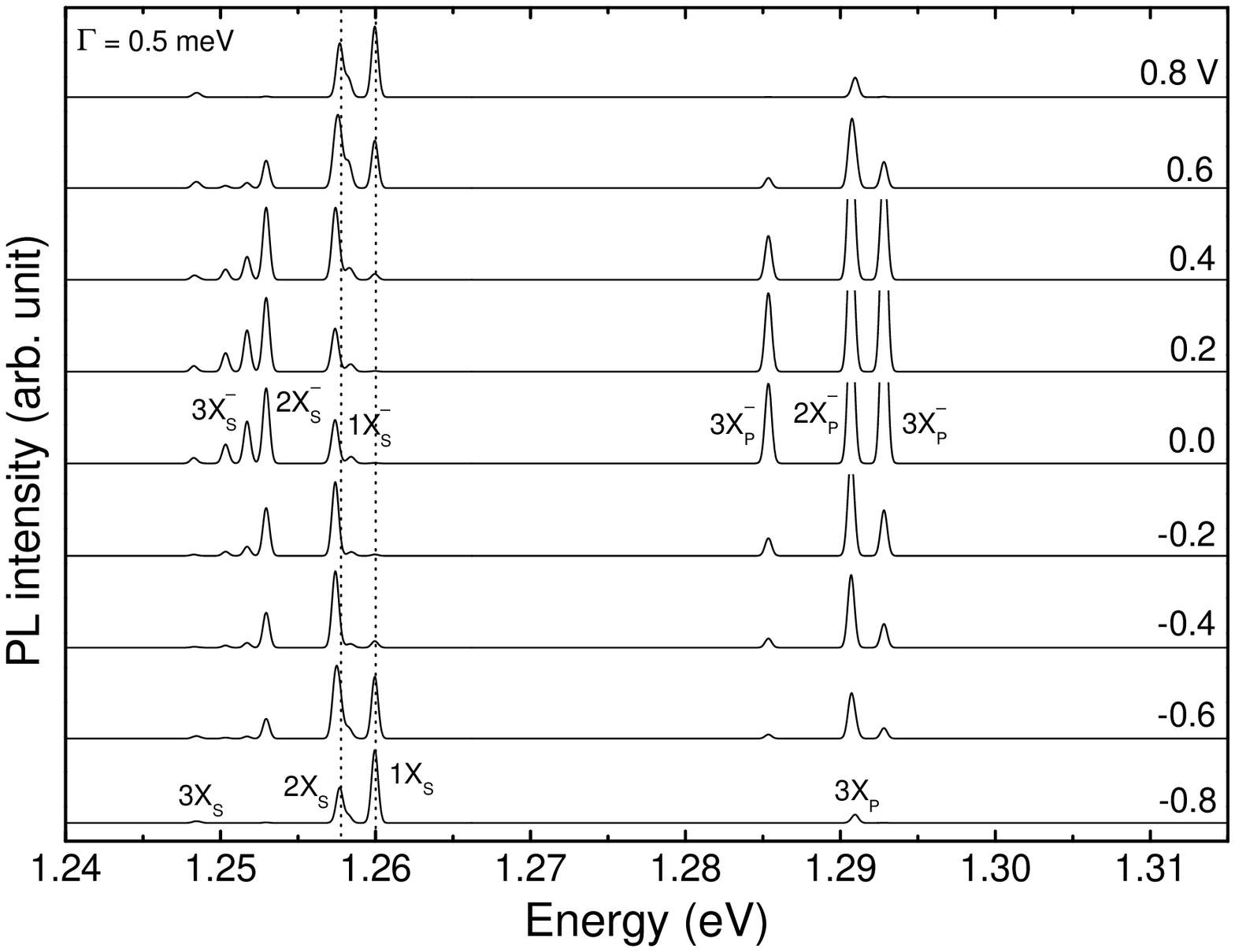,width=15cm}}
\caption{Calculated PL spectra of a coupled QD molecule with inter-dot distance
$d\!=\!4.5$ nm at different applied voltages. The spectra lines are artificially 
broadened by 0.5 meV.}
\label{fig4}
\end{figure}

In Fig.\ref{fig4} we display the calculated spectra at different applied voltages (we
assume a linear drop of the voltage across the intrinsic layer of the sample). The
main peaks in the extreme voltages spectra ($\!\pm 0.8$ V) are related to several 
possible optical transitions from the various neutral multi-exciton configurations 
\cite{ref17,ref20}, in QD1: $1X_S$ (1.2599 eV), $2X_S$ (1.2577 eV), $3X_S$ (1.2484 eV, 
1.2582 eV), $3X_P$ (1.291 eV). The spectrum at 0 V is dominated by configurations of negatively charged multi-excitons: $1X_S^-$ (1.2574 eV), $2X_S^-$ (1.2529 eV), $2X_P^-$ (1.2907 eV), 
$3X_S^-$ (1.2483 eV, 1.2503 eV, 1.2517 eV), $3X_P^-$ (1.2853 eV, 1.2928 eV). The calculated 
spectra reproduce well the main features of the experimental spectra (Fig.\ref{fig2}b). 
A similar calculation, which does not take into account charge transfer between the dots, 
leads to a qualitatively different spectral behavior \cite{ref19}.

In summary, we have measured the time-averaged PL of single vertically stacked InAs QD
pairs under applied electric field. For the closely spaced QD pair (molecule) we find
that coupling between excited states of the two dots leads to field tunable electron
transfer from one dot to the other. Our theoretical model reproduces the experimental
results by simultaneously taking into account the field-dependent carriers kinetics in
the QD molecule, and the molecule's many-body spectrum for different carriers populations.

This work was supported by the ARO-DARPA grant and the Humboldt Foundation.

\newpage

\onecolumn

\newpage

\newpage

\newpage

\begin{references}
\bibitem{ref15} The PL intensity ratio QD1/QD2 varies for different samples with
                nominally the same structure, and may be related to some extrinsic,
                non-radiative recombination channels.
\end{references}
\end{document}